# Control of topography, stress and diffusion at molecule-metal interface.

**Nikolai B. Zhitenev**[1,5], **Weirong Jiang**[1,2], **Artur Erbe**[3], **Zhenan Bao**[4], **Eric Garfunkel**[2], **Donald M. Tennant**[1] **and Raymond A. Cirelli**[1]

[1] Bell Labs., Lucent Technologies, Murray Hill, New Jersey 07974

[2] Department of Chemistry, Rutgers University, Piscataway, New Jersey 08854

[3] Fachbereich Physik, Universität Konstanz, 78464 Konstanz, Germany

[4] Department of Chemical Engineering, Stanford University, Stanford, California 94305

email: zhiten@lucent.com

**Abstract.**

Transport properties of metal-molecule-metal junctions containing monolayer of conjugated and saturated molecules with characteristic dimensions in the range of 30-300 nm are correlated with microscopic topography, stress and chemical bonding at metal-molecule interfaces. Our statistically significant dataset allows us to conclude that the conductivity of organic molecules ~1.5 nm long is at least 4 orders of magnitude lower than is commonly believed.

---

[5] Corresponding author

PACS: 73.20.-r, 73.22.-f, 73.40.Gk, 73.40.Rw, 73.63.Rt

**1. Introduction**

Reliable and scalable integration of organic molecules within nanoscale electronic devices has the potential to dramatically expand available device functionality. Similar to other device platforms such as Si-based technology, the electronic properties of devices that are just a few atomic layers thick are determined not solely by the properties of the host material but are equally dependent on dopants, defects and electronic states at interfaces. Incorporation of molecules in small devices calls for the simultaneous solutions of many interrelated material, electronic and chemical issues. In this paper, we focus on devices built of self-assembled molecular layers (SAM) enclosed between two metal electrodes[1-3]. Typically, organic molecules do not have electronic states at the energy close to the Fermi energy of common metals. Electrons travel between metal electrodes by tunneling through the molecules in the gap between highest occupied (HOMO) and lowest unoccupied (LUMO) molecular orbitals. Tunneling transmission is governed by the energy difference between the Fermi level of the contacts and the closest molecular orbital and by the spatial extent of the molecular orbital. Because of the exponential sensitivity of tunneling transmission to energy and distance, the conductance of real devices can be much larger or much smaller than that from molecular tunneling.

A single state in a molecular layer with defect energy close to the Fermi energy can dramatically increase the transmission. The most common origin of defects in molecular devices is penetration of metal contacts into the molecular layer. During and after fabrication, metal filaments can form, either completely shorting the source and the drain electrodes and limiting the yield of useful devices[4, 5], or strongly increasing device conductance[6]. Metal particles can penetrate into the molecular layers[7]. In the process of the metal contact film growth and crystallization the electrode can protrude into the layer, thus deforming molecules and affecting conductance[8]. On the other hand, distorted chemical bonds at the interface between metal and molecule can decrease the tunneling transmission. Metal films commonly used for electrodes, e.g., Au, Ag, Ti, Pt grow and interact very differently on top of molecular layers[9-12]. Currently, every experiment is unique in that the atomic placement of the relevant device constituents cannot be determined accurately.

The main goal of the paper is to study the phenomena affecting the conductance of molecular devices by systematically varying the growth conditions at the metal-molecule interface. Specifically, first we optimize the surface topography of the metal electrode used for the assembly of molecular layer reducing the density of structural defects in the SAM. This dramatically reduces the diffusion of the top metal contact through the molecular layer, increasing the device yield to >90%. Then, we experiment with chemical bonding and surface topography at the top metal-molecule interface.

2. Experimental details

The general approach for molecular device fabrication is to perform the most critical patterning of nanometer features without molecules, assemble the molecules, and to complete the structure with a relatively non-invasive processing step. We use small shadow masks defined within a stack of $SiO_2/SiN_x/SiO_2$ layers grown on degenerately doped Si substrates to obtain features below the lithographic limit[5, 13-16]. The fabrication of metal-SAM-metal junctions using the masks is illustrated in Figure 1. First, bottom electrodes are defined by evaporation. In the current study, the bottom electrode is Ti/Au (5A/300A). The electrodes are separated by a bridge with a width in the range of 100-300 nm. Next, a SAM is deposited from solution in the usual manner. Two types of molecules, representing opposite ends of expected electronic functionality, are used in this study. Terthiophenedithiol (T3), synthesized using previously described methods[17], is a conjugated molecule with thiol groups responsible for chemical attachment to metal electrodes. Decanedithiol (C10) is a fully saturated molecule with a length of 1.5 nm similar to T3 length. The substrates with bottom Au electrodes were soaked in a tetrahydrofuran (THF) solution of the thiols (about 0.01 mM) at room temperature for 24 hours, then rinsed with THF, toluene and isopropanol. Both molecules form SAMs with thiol terminations exposed at the top interfaces as proven by nanotransfer experiments[18]. Finally, a top electrode is evaporated through the same mask from a different angle. The size of the junction is controlled by the size of the bridge and the angle of the second evaporation. SEM images of representative junctions are shown in Figure 1c. A single chip contains 84 separate devices, allowing us to perform statistical analysis on nominally equivalent junctions and to vary the junction sizes. The electrical characterization does not require the removal of the mask stack or metal accumulated on top surface of the mask, thus minimizing this potential source of damage or contamination.

3. Results.

In the first experiments, we assembled both molecules on an as-deposited Au electrode. The top electrode was an 8 nm Au or Ag film. Resistance of all junctions appeared to be indistinguishable from the leads resistance which is in the range of 300 Ω-1 kΩ for devices with different lead geometries. We conclude that the evaporated metal penetrates through the SAMs shorting the devices. These results are consistent with our earlier observation of a low yield of non-shorted multi-grain junctions formed with different nano-templates[19]. The evaporated top metal easily diffuses near defects in molecular packing induced by multiple grain boundaries of the bottom electrode.

In the next set, we anneal (250 C, 5 min) the chips after deposition of the bottom Au contact. The annealing modifies the grain structure of polycrystalline Au film, making the grains smother and the grain size larger. The distribution of device resistances measured at room temperature is shown in Fig. 2a. Although the variation is very broad, the yield of non-shorted devices is above ~96% for T3 and ~92 % for C10 SAMs; this is much higher than in previous experiments by others (0.5% - 5%) with similar electrode arrangements[4, 5, 15]. Surprisingly, the apparent median resistance for T3 devices is higher than for C10 devices, contrary to all expectations. This clearly shows that the conductance of real junctions is not directly determined by the electronic structures of the molecules. No scaling of the resistance with the junction area is observed.

To study bond formation at the top interface, we used overlayers of metals with different chemical reactivities. Au, Ag or Ti (7nm) were studied (Figure 2). Silver is more reactive than Au, and titanium is known to strongly react not only with thiol groups at the top interface but also with carbon atoms[12]. The electrical properties of T3-Ag junctions are generally similar to those of T3-Au junctions. A significant percentage of the C10-Ag junctions were shorted. The conductance of the junctions with Ti overlayers is consistently higher. The overall histograms are very similar for both T3 and C10 SAMs.

The microscopic topography at the top SAM-metal interface is generally unknown. In an attempt to control the topography, we intentionally create clusters at the top interface. First, 0.3-0.5 nm of Au is evaporated on top of the SAM, simultaneously on the chips and on reference SAMs assembled on atomically flat Au substrate. The evaporation chamber is vented, and the reference sample is used to examine the topography of the overlayer by STM. Clusters with average diameter ~6 nm are clearly seen (Figire 3f). The junction fabrication is completed by evaporating an 8nm thick film of Au. The electrical properties of the junctions changes dramatically in comparison with the uninterrupted Au evaporation as shown in Figure 3a and 3b. All T3 junctions are shorted while all C10 junctions are highly resistive (for convenience, we lump all $R>10^{12}\Omega$ in a single bin).

In the next experiment, after deposition of 0.5 nm of Au, the evaporation is interrupted for 3 min followed by continuous evaporation of 8nm of Au (Figure 3c and 3d). The results clearly fall between the continuous Au evaporation shown in Figure 2 and the previous experiment. Half of the T3 junctions are not shorted, and the C10 junctions are less resistive on average. The shorted T3 junctions can be electrically driven into a more resistive state. Voltage pulse with ~30-50 mV amplitude and rise time below 1 μs usually triggers modification of the shorted junctions. We note that applying dc voltage up to 1V does not change the junction conductance. The distribution of resistances of junctions after the breakdown is shown in Figure 3e.

Junction resistance measured at room temperature only partly characterizes the transport properties. We studied conductance of the junctions as a function of source-drain voltage and temperature. The specific details of current-voltage (I-V) curves vary significantly as can be expected from the broad distribution of conductance values. Some representative results are illustrated in Figure 4.

Transport characteristics of all T3-Ti and C10-Ti junctions are quite similar (Figure 4a). Conductance falls only by 5-15% as temperature changes from 300 K to 4.2 K. Low-temperature dI/dV curves are rather smooth, with a small dip near zero. The lower conductance, the zero-bias dip and the smooth dI/dV variation clearly differentiate these junctions from shorted ones. Weak temperature dependence and zero-bias anomaly are the usual signatures of tunneling conductance. Most of the non-shorted C10-Ag and all C10-Au$_{clust}$ junctions also display insignificant temperature dependence but the tunneling conductance in this case is lower by orders of magnitude in comparison with Ti overlayer junctions. A small percentage of other types of junctions also display weak temperature dependence of conductance.

The transport properties of majority of other junctions with Au and Ag overlayers are more complicated. I-V curves are usually linear at room temperature. Conductance measured at zero source-drain bias falls noticeably with decreasing temperature (Figure 4b). Low-temperature I-V curves display a non-linear region with characteristic voltage scale $V_{sd}$ ~ 25-200 mV. This general behavior is typical for T3-Au, C10-Au, T3-Ag, T3-Au$_{clust}$ junctions. Similar behavior has been observed in our previous study of multi-grain junctions fabricated on the tips[19]. Two separate conductance mechanisms contribute to the overall electrical transport. The temperature dependent part of the conductance can be identified as hopping transport, characterized by a small energy scale in the range ~10-150 meV. The residual conductance observed at low temperature is a combination of direct tunneling between contacts, sequential tunneling through low-energy defect states and hopping. The relative contributions of hopping and tunneling to the overall conductance vary broadly from sample to sample.

Finer differences between transport behavior in T3 and C10 SAMs can be seen in the low-temperature dI/dV curves. Conductance peaks shown in Fig. 4b are observed in most of the T3-Au, T3-Ag and T3-Au$_{clust}$ junctions. The positions of the peaks on dI/dV($V_{sd}$) curves can be shifted by gate voltage in the case of T3-Au and T3-Ag junctions. Such behavior is reminiscent of single-electron charging of isolated islands. In T3-Au$_{clust}$ junctions, the peaks are stronger relative to the smoother background while the peak positions are usually insensitive to the gate voltage. No similar conductance peaks were observed in the majority of C10-Au samples.

## 4. Discussion

First, we comment on overall conductance values through the SAM. Commonly accepted[20] tunneling conductance per conjugated molecule of comparable length[21, 22] is $10^{-6}$-$10^{-8}$ $\Omega^{-1}$ and conductance per alkane molecule[23] is $10^{-8}$-$10^{-9}$ $\Omega^{-1}$. A median junction ~100 nm by 100 nm contains ~$5*10^4$ molecules. If one assumes that every molecule is well-bonded on both sides, the median resistance of T3 junctions is 20 Ω – 2 kΩ and 2 – 20 kΩ for C10 junctions. If we conservatively assume having just a single bond per metal cluster of the top contact, a representative junction contains ~300 well-bonded molecules. The corresponding resistance estimates are 3 kΩ- 300 kΩ for T3 junctions, and 300 kΩ – 3 MΩ for C10 junctions. Clearly, the results show that the tunneling conductance of molecules is lower by 4-6 orders of magnitude for both conjugated and saturated molecules, in strong contrast to the majority of previous calculations and experiments[20].

In fact, in the vast majority of devices we cannot single out the conductance associated with tunneling through the molecular orbitals. High values of tunneling conductance measured in the junctions with Ti overlayers are unlikely related to the electronic structures of the original molecules since no difference between conjugated and saturated molecules is seen in the experiment. Evaporated Ti reacts strongly with organics[12] attacking the SAM and forming Ti carbides and oxycarbides. Tunneling conductance measured in all other junctions is likely to be determined by the microscopic configuration of metal electrodes that have partly penetrated the SAM, and/or a small number of defect states rather than by molecular states.

The essential material transformations defining the electronic properties of molecular junctions are schematically illustrated in Figure 4c. Structural defects in SAMs assembled on as-deposited Au allow for easy diffusion of the top metal electrode through the layer. The SAM structure is significantly improved by annealing Au films before the deposition of molecules. The high yield of non-shorted devices achieved by evaporation of Au on top of SAMs suggests that practically all incident Au atoms are stopped at the top interface. The overall electrical properties of junctions formed with T3 and C10 SAMs are rather similar. We believe that during its growth and crystallization the Au overlayer penetrates deep into the SAM, exerting substantial pressure and deforming the molecules. The microscopic details of the electrode topography are almost independent of the molecule type in this particular case. Diffusion and growth of Ag overlayers differ depending on the SAM type. Ag penetrates alkane SAMs more easily. Finally, evaporated

Ti can strongly react with SAMs, modifying electronic structure with little discrimination between conjugated and saturated molecules.

Relaxation of the thin Au overlayer accumulated at the top molecular interface and cluster formation results in two new phenomena. First, comparing properties of the junctions formed on C10 SAMs, we suggest that this relaxation significantly reduces stress at the interface and the penetration of the top metal contact into the SAM volume. The C10 junctions with a relaxed top interface are so resistive that their conductance cannot be reliably differentiated from a possible leakage through the substrate. Very different behavior is seen on conjugated SAMs. Effectively, the clusters diffuse through the SAM easier than separate Au atoms impinging the SAM during the evaporation. The cluster diffusion has to proceed along with a redistribution of molecules in the SAM. Apparently, the configuration with the clusters partly or fully submerged into conjugated SAM lowers the total energy of the system. The clusters that form at the edge of a continuous Au top electrode and partly penetrate into the SAM can account for the systematic observation of conductance resonances sensitive to the gate voltage in T3 junctions.

The results clearly expose a variety of material transformations and self-organization processes occurring during integration of organic and inorganic components in nanoscale devices. This is the first experimental research that systematically correlates the electrical properties of SAM-based molecular devices with the microscopic details of the metal-molecule interface. We have demonstrated that the generation of defects can be dramatically reduced for certain combinations of metals and molecules by changing surface topography and growth conditions at the interface.

**Acknowledgements**

**Acknowledgements**

We would like to acknowledge useful discussions with E. Chandross and D. Hamann.

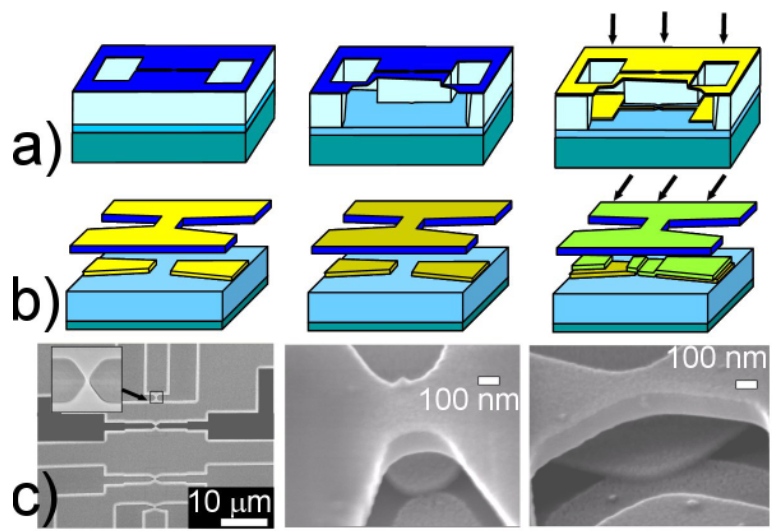

**Figure 1.** Fabrication of templates and molecular junctions. a). The stencil mask is defined within an insulating stack of $SiO_2/Si_3N_4$ layers grown on Si substrate. A layer of $SiO_2$ (light blue, 10 nm in 5 μm by 5 μm central windows, 200 nm elsewhere on the wafer) is grown on Si (light purple) to isolate the substrate from the devices followed by $Si_3N_4$ (~400 nm, cyan) and $SiO_2$ (~150 nm, blue) mask forming layers. The desired pattern is defined by photolithography and etching of the top $SiO_2$ layer. $Si_3N_4$ is selectively etched undercutting the top $SiO_2$. The mask is used to define the pattern of evaporated metal electrodes (yellow). The metal accumulated on the top surface of the mask is electrically isolated from the evaporated electrodes. b) Fabrication of molecular junction using the template. Bottom electrodes are defined on the substrate by evaporation. Molecular layer is deposited from solution. Evaporation of second electrode from an angle completes the junction. c) SEM images of the template (one out of four shown samples is shaded) and examples of small and large junctions.

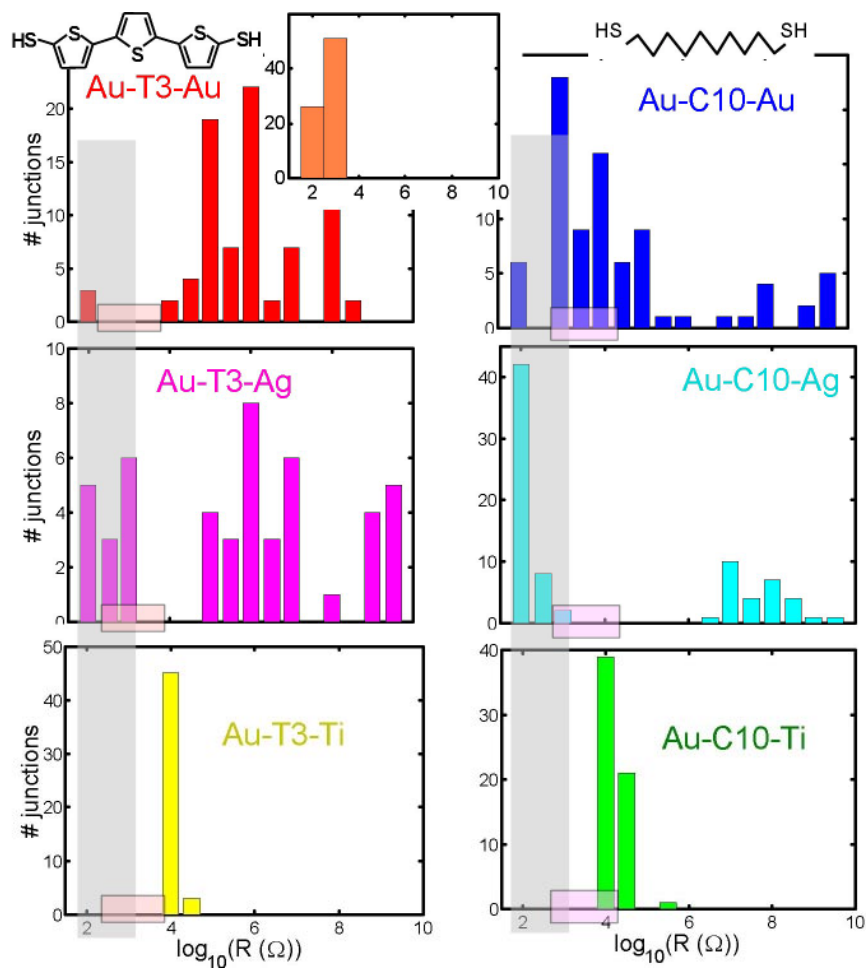

**Figure 2.** Distribution of junction resistances measured at room temperature. Resistance range corresponding to the leads resistance is gray colored (shorted junctions). Light-colored bars show expected ranges of resistances assuming that (i) all molecules inside junctions are bonded to both electrodes and (ii) tunneling conductance of $10^{-6}$-$10^{-8}$ $\Omega^{-1}$/T3 and $10^{-8}$-$10^{-9}$ $\Omega^{-1}$/C10. Inset in the top row shows results for junctions fabricated on as-deposited Au electrode.

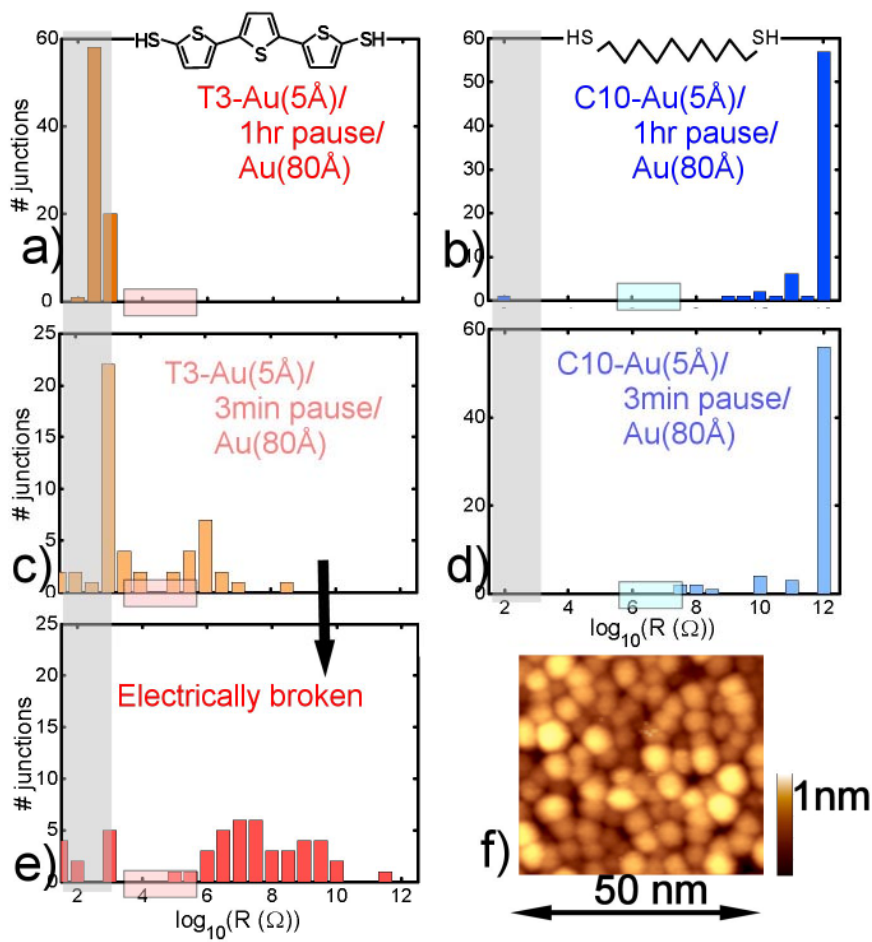

**Figure 3.** Distribution of junction resistances with the clusters created at the top metal-molecule interface. Top metal contact is formed by depositing 0.5 nm of Au film first, letting the film relax for time $\tau_{rel}$, and depositing another 8 nm of Au: a) $\tau_{rel}$=1 hr, T3 SAM b) $\tau_{rel}$=1 hr, C10 SAM c) $\tau_{rel}$=3 min, T3 d) $\tau_{rel}$=3 min, C10. e) distribution of junction resistance from (c) after the soft electrical breakdown. Colored bars show expected range of resistances assuming that (i) only one molecule per cluster forms good bonds at both ends and (ii) tunneling conductance of $10^{-6}$-$10^{-8}$ $\Omega^{-1}$/ T3 and $10^{-8}$-$10^{-9}$ $\Omega^{-1}$/C10. f) STM image of clusters formed on top of T3 SAM.

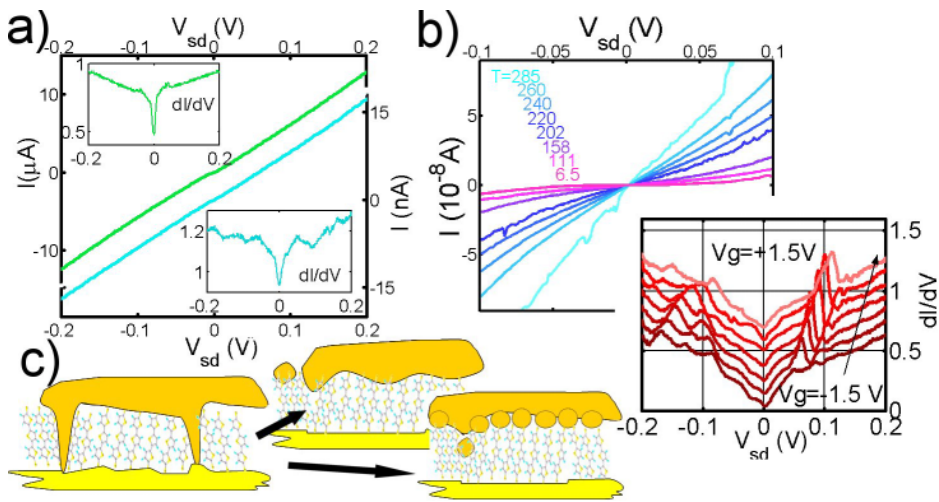

**Figure 4**. Examples of characteristic transport behavior. a) IV curves measured on C10-Ti (top) and C10-Ag (bottom) junctions. The current scales are different. T= 8 K. Insets: Corresponding differential conductance dI/dV as a function of $V_{sd}$ b) IV curves measured at different temperatures on C10-Au junction. Inset: Set of dI/dV curves measured on T3-Au junction at different gate voltages. T=8 K. c) Schemes of metal assembly and penetration into molecular layer illustrating material modifications and interface topography studied in this paper. Junctions assembled on rough Au surface are usually shorted because of easy metal diffusion along structural defects of the SAM. Yield of non-shorted junctions is high if bottom Au electrode is annealed before the SAM deposition. Metal electrodes can protrude into the SAM volume and/or deform molecules. Relaxation of thin metal underlayer reduces stress at the interface.